\documentstyle[12pt]{article}

\def\[{\left\lbrack}
\def\]{\right\rbrack}

\def\({\left(}
\def\){\right)}
\def\ih{\'\i}

\title{Operatorial quantization of Born-Infeld Skyrmions}

\author{J. Ananias Neto\thanks{e-mail:jorge@fisica.ufjf.br},
W. Oliveira\thanks{e-mail:wilson@fisica.ufjf.br}
and Emanuel R. de Oliveira\thanks{e-mail:emanuel@fisica.ufjf.br}\\
Departamento de F\ih sica, ICE \\ Universidade Federal de Juiz de
Fora, 36036-330, \\ Juiz de Fora, MG, Brazil }

\date{}

\begin{document}

\maketitle

\begin{abstract}
\noindent The SU(2) collective coordinates quantization of the 
Born-Infeld\break Skyrmions Lagrangean is performed. 
The obtainment of the classical Hamiltonian from 
this special Lagrangean is made by using an approximate way:
it is derived from the expansion of this non-polynomial Lagrangean 
up to second-order variable in the collective coordinates, 
using for this some causality arguments. Because this 
system presents constraints, we use the Dirac Hamiltonian method 
and the Faddeev-Jackiw Lagrangean approach to quantize this model. 

\end{abstract}

\noindent PACS number: 11.10.Ef; 12.39.Dc\\
Keywords: Skyrme model, constrained systems.
\maketitle

\setlength{\baselineskip} {20 pt}
\newpage

   Until this moment, the Skyrme model\cite{Skyrme} is the 
effective field theory for baryons and their interactions. These 
hadronic particles are described from soliton solutions in the non-linear 
sigma model. In this Lagrangean, normally, it is necessary to add the 
Skyrme term to stabilize the soliton solutions. The physical spectrum 
is obtained by performing the collective coordinates quantization. Then,
using the nucleon and the delta masses as input parameters, we get 
the principal phenomenological results\cite{Adkins}.

In principle, the Skyrme term is arbitrary, having no
concrete reason to fix this particular choice\cite{Weinberg}. Its 
importance resides in the fact that, maybe, it is the most simple possible
quartic derivative term that we need to put in the static Hamiltonian to obtain 
the soliton solution\footnote{ According to Derrick scale
theorem\cite{Derrick}.}. 
However, it is possible to get round this ambiguity by adopting a 
non-conventional Lagrangean also based in a non-linear sigma model, 
given by

\begin{equation}
\label{BF}
L= - F_\pi \int d^3r \, Tr \[ \partial_\mu U 
\partial^\mu U^+ \]^{3\over 2},
\end{equation}

\noindent where $F_\pi$ is the pion decay constant and $U$ is an
SU(2) matrix. This model was proposed by  Deser, Duff and
Isham\cite{Deser}, based on the ideas of Born-Infeld Electrodynamics\cite{BInf}. 
Applying Derrick theorem in the static Hamiltonian (derived from the Lagrangean (\ref{BF}) ) we observe the existence of soliton solution.

The purpose of this paper is to apply both Dirac Hamiltonian method\cite{Dirac}
and the Faddeev-Jackiw Lagrangean procedure\cite{Faddeev} to obtain the commutators
between the canonical operators in which will define the quantum structure
of the Born-Infeld Skyrmions system\cite{Marleau}. We will observe
that when we keep the non-causality sector of the soliton solution
influencing the physical values as minimum as possible, the commutators
obtained are the same of the Skyrme model.

The dynamic system will be given by performing the SU(2) collective 
semi-classical expansion\cite{Adkins}. Substituting $U(r,t)$ by 
$A(t)U(r)A^+(t)$ in (\ref{BF}), where $A$ is an SU(2) 
matrix, we obtain

\begin{equation}
\label{BFC}
L = - F_\pi \int d^3r \[ m - I\, Tr ( \partial_0 A 
\partial_0 A^{-1} ) \]^{3\over 2}.
\end{equation}

\noindent In the last equation, $m$ and $I$ are
some functional of the chiral angle $F(r)$, with the topological
boundary conditions,$\,F(0)=\pi$ and $F(\infty)=0$. Here,
we use the Hedgehog ansatz for $U$, i.e., $U=\exp(i\tau\cdot \hat{r}
F(r))$. The SU(2) matrix 
A can be written as $A=a_0+i a\cdot \tau$, with the constraint

\begin{equation}
\label{primary}
\sum_{i=0}^{i=3} a_ia_i = 1.
\end{equation}

\noindent The Lagrangean (\ref{BFC}) can be written as a function
of the $a_i$ as

\begin{equation}
\label{BFCOL}
L = - F_\pi  \int d^3r \[  m - I \, \dot{a}_i\dot{a}_i  \]
^{3\over 2}.
\end{equation}

\noindent From the Eq. (\ref{BFCOL}) we can obtain the conjugate momentum given by

\begin{eqnarray}
\label{momen}
\pi_i = {\partial L \over \partial \dot{a}_i }
= 3 F_\pi \,\,\dot{a}_i  
\int d^3r I \[ m -  I\,\dot{a}_k\dot{a}_k \]^{1 \over 2}.
\end{eqnarray}

\noindent The algebraic expression for the Hamiltonian is 
obtained by applying the Legendre transformation, 
$ \, H = \pi_i\dot{a}_i - L \,$. 
However, due to the momentum formula given in Eq.~(\ref{momen}), 
maybe it is not possible to write the conjugate Hamiltonian in 
terms of $\pi_i$
and $a_i$ for the Lagrangean of the Born-Infeld Skyrmions. 
An alternative procedure is to expand the original Lagrangean 
(\ref{BFCOL}) in collective 
coordinates variables. Thus, considering the binomial expansion variable
$\,\,{I\over m} \,\dot{a}_i\dot{a}_i \,\,$, the Lagrangean sum is given by

\begin{eqnarray}
\label{Lseries}
L = - M + A (\dot{a}_i\dot{a}_i) - B ({\dot{a}_i\dot{a}_i}^2)
+ \dots \,,
\end{eqnarray}

\noindent where

\begin{eqnarray}
\label{definitions1}
M &= &F_\pi \int d^3r \,\, m^{3\over 2},\\
\label{definitions2}
A &=& 3F_\pi \int d^3r \,\, I \sqrt{m},\\
\label{definitions3}
B &= &{3\over 2} F_\pi \int d^3r \,\, {I^2 \over \sqrt{m}},\\
\vdots
\end{eqnarray}

In this step we would like to give a physical
argument that permits the use of this procedure. This model is not 
relativistic invariant. Thus, we hope that only soliton velocity 
much smaller than speed of light can reproduce, with a good accordance, the experimental physical results. From the relation\cite{Weigel}, $A^+ {\partial_0 A}= i/2\sum_{k=1}^{k=3} \tau_k \omega_k$, where $\omega_k$ is 
the uniform soliton angular velocity, it is possible to show that
$Tr[\partial_0 A \partial_0 A^+] = 2 \dot{a_i}\dot{a_i}=\omega^2/
2$. If we want that the soliton rotates with velocity smaller than {\it c},
then  $\omega r \ll 1$, leading to $\dot{a_i}\dot{a_i}={\omega^2\over 4} \ll 1$, and consequently
$\dot{a_i}\dot{a_i} \ll 1$ for all space\footnote{In the context of semi-classical expansion, it is expected that the product of $\dot{a_i}\dot{a_i}$ by the expression ${I\over m}$ given by the Euler-Lagrange equation, does not modify sensitively this result.}. Thus, these results explain our procedure.

The Hamiltonian is obtained by using the Legendre transformation

\begin{eqnarray}
\label{Hamil1}
H & = & \pi_i \dot{a}_i - L \nonumber \\
& = & M + A (\dot{a}_i\dot{a}_i) -3B {(\dot{a}_i\dot{a}_i)}^2 + \dots\, .
\end{eqnarray}

\noindent Obtaining the momentum,$\,\pi_i={\partial L\over 
\partial \dot{a_i}} \,$, from Eq. (\ref{Lseries}), writing the 
the Lagrangean as, $L=\pi_i\dot{a_i}-H$, and comparing with the 
expansion of the Lagrangean (\ref{Lseries}), it is possible to 
derive the expression of the Hamiltonian (\ref{Hamil1}) as

\begin{eqnarray}
\label{Hamilf}
H =  M + \alpha\, \pi_i\pi_i + \beta \, {(\pi_i\pi_i)}^2
 + \dots \,,
\end{eqnarray}

\noindent being $\, \alpha={1\over 4A} ,\beta={B\over 16A^4}\,$
($\,\,A\,$ and $\,B\,$ are defined in Eqs.(\ref{definitions2}) and 
(\ref{definitions3}) respectively). We will truncate the 
expression (\ref{Hamilf}) in the second order variable\footnote{
Due to the equation(\ref{momen}) together with the fact that 
$\dot{a_i}\dot{a_i} \ll 1\,\,$, we expect that terms like 
${(\pi_i\pi_i)}^3$ or higher order degree do not alter our
conclusion about the commutators of the quantum Born-Infeld Skyrmions.}
, and we will use this approximate Hamiltonian to perform the 
quantization.

In order to apply the Dirac Hamiltonian method, we need to look for 
secondary constraints, in which can be calculate by constructing the
following Hamiltonian

\begin{eqnarray}
\label{formula1}
H_T &=& H + \lambda_1 \phi_1 \nonumber \\
&=& M+ \alpha \, \pi_i \pi_i + \beta \,(\pi_i\pi_i)^2 + 
\lambda_1 (a_ia_i-1),
\end{eqnarray}

\noindent where $\,\lambda_1\,$ is the Lagrangean multiplier. 
Imposing that the primary constraint\\(\ref{primary}) must be 
conserved in time, we have

\begin{eqnarray}
\label{conserved}
\dot{\phi_1} = \{ \phi_1,H_T \} =0,
\end{eqnarray}

\noindent where $\,\phi_1=a_ia_i-1\approx 0\,$ is the primary constraint.
The consistency condition over the constraint (\ref{primary}) 
leads to

\begin{eqnarray}
\label{time2}
\dot{\phi_1}=\{ \phi_1, H_T \}
=\{a_ia_i-1, M + \alpha \pi_j\pi_j + \beta {(\pi_j\pi_j)}^2 \}\nonumber\\
=(4\alpha + 8\beta \pi_j\pi_j ) a_i\pi_i.
\end{eqnarray}

\noindent As the quantity inside the parenthesis, formula (\ref{time2}),
 can not be a negative value, then we get the secondary constraint

\begin{eqnarray}
\label{secondary}
a_i\pi_i = 0, \Rightarrow \phi_2=a_i\pi_i\approx 0.
\end{eqnarray}

\noindent If one goes on and imposing the consistency condition over (\ref{secondary}), we observe that no new constraints are obtained 
via this iterative procedure. The theory has then the constraints

\begin{eqnarray}
\label{formula3}
\phi_1 &=& a_ia_i - 1 \approx 0, \\
\label{formula4}
\phi_2 &=& a_i\pi_i \approx 0,
\end{eqnarray}

\noindent which are second class ones. To implement the Dirac brackets
we need to calculate the matrix elements of their Poisson brackets, read
as

\begin{eqnarray}
\label{formula5}
C_{\alpha\beta} &=& \{\phi_\alpha,\phi_\beta\} \\
&=& -2\epsilon_{\alpha\beta}\,a_ia_i, \,\,\,\,\,\, \alpha,\beta=1,2,
\end{eqnarray}

\noindent with $\epsilon_{12} = -\epsilon^{12} = -1$. Using the Dirac 
bracket

\begin{equation}
\label{formula 6}
\{A,B \}^*= \{A,B \}-\{A,\phi_\alpha \}C^{-1}_{\alpha\beta}
\{\phi_\beta,B \},
\end{equation}

\noindent we obtain 

\begin{eqnarray}
\label{formula 7}
\{ a_i,a_j \}& = & 0,\\
\label{formula 8}
\{ a_i,\pi_j \}& = & \delta_{ij} - a_i a_j ,\\
\label{formula 9}
\{ \pi_i,\pi_j \}& = &  a_j\pi_i - a_i\pi_j.
\end{eqnarray}

\noindent By means of the well known canonical quantization rule
$\{\,,\,\}^* \rightarrow -i\[\,,\,\]$, we get the commutators

\begin{eqnarray}
\label{formula10}
\[ a_i,a_j \]& = & 0,\\
\[ a_i,\pi_j \]& = & -i \( \delta_{ij} - a_i a_j \) ,\\
\[ \pi_i,\pi_j \]& = & -i \( a_j\pi_i - a_i\pi_j \).
\end{eqnarray}

\noindent These results show that the quantum mechanics commutators 
of the Born-Infeld Skyrmions, when we expand its Lagrangean until to
the second order in collective coordinates, are equal to the Skyrme model\cite{Fujii}.

To implement the Faddeev-Jackiw quantization procedure\cite{Barcelos}, 
let us consider the first-order Lagrangean

\begin{equation}
\label{formula11}
L = \pi_i\dot{a}_i - V,
\end{equation}

\noindent where the potential $V$ is

\begin{equation}
\label{formula12}
V = M + \alpha\pi_i\pi_i + \beta(\pi_i\pi_i)^2 + \lambda (a_ia_i - 1),
\end{equation}

\noindent with the sympletic enlarged variables given by
$\xi_j=(a_j,\pi_j,\lambda)$. To obtain the Poisson brackets, we 
need to determine the sympletic tensor, defined by

\begin{eqnarray}
\label{tensor}
f_{ij} = {\partial A_j\over \partial \xi^i}
-{\partial A_i\over \partial \xi^j}.
\end{eqnarray}

\noindent $A_i$ are functions of the sympletic variable 
$\,\xi\,$. From the Lagrangean(\ref{formula11}), we identify the 
coefficients

\begin{eqnarray}
\label{formula13}
A_{a_i} &=& \pi_i, \nonumber \\
A_{\pi_i} &=& 0, \nonumber \\
A_{\lambda} &=& 0,
\end{eqnarray}

\noindent The sympletic tensor is calculate using the 
definition (\ref{tensor}) with the coefficients above leading to the
matrix $f$

\begin{equation}
f = \left(
\begin{array}{ccc}
0           & -\delta_{ij} & 0 \\
\delta_{ij} &         0     & 0 \\
0           &         0     & 0
\end{array}
\right)
\end{equation}

\noindent where the elements of rows and columns follow the order: 
$a_i$, $\pi_i$, $\lambda$.
The matrix above is obviously singular. Thus, the system has 
constraints in the Faddeev-Jackiw formalism. The eigenvector 
corresponding to the zero eigenvalue is

\begin{equation}
v = \left(
\begin{array}{ccc}
0 \\
0 \\
1
\end{array}
\right)
\end{equation}

\noindent The primary constraint is obtained from

\begin{eqnarray}
\label{formula15}
\Omega^{(1)} &=& v_i \frac{\partial V}{\partial A_i} \nonumber \\
&=& a_ia_i -1 \approx 0,
\end{eqnarray}

\noindent where the potential $V$ is given by Eq.~(\ref{formula12}).
Taking the time derivative of this constraint and 
introducing the result into the previous Lagrangean by means 
of a Lagrange multiplier $\rho$, we get a new Lagrangean $L^{(1)}$

\begin{equation}
\label{formula16}
L^{(1)} = (\pi_i + \rho a_i)\dot{a}_i - V^{(1)},
\end{equation}

\noindent where

\begin{equation}
\label{formula17}
V^{(1)} = M + \alpha\pi_i\pi_i + \beta (\pi_i\pi_i)^2.
\end{equation}

\noindent The new coefficients are\footnote{We have imposed strongly
$a_ia_i-1=0.$}

\begin{eqnarray}
\label{formula18}
A_{a_i}^{(1)} &=& \pi_i + \rho a_i, \nonumber \\
A_{\pi_i}^{(1)} &=& 0, \nonumber \\
A_{\rho}^{(1)} &=& 0.
\end{eqnarray}

\noindent The matrix $f^{(1)}$ is then 

\begin{equation}
f^{(1)}=\left(
\begin{array}{ccc}
0           & -\delta_{ij} & -a_i \\
\delta_{ij} &         0     & 0 \\
a_i           &         0     & 0
\end{array}
\right)
\end{equation}

\noindent where rows and columns follow the order: $a_i$, $\pi_i$, $\rho$. 
The matrix $f^{(1)}$ is singular. An eigenvector corresponding to 
the zero eigenvalue is

\begin{equation}
v^{(1)}=\left(
\begin{array}{ccc}
0 \\
a_i \\
-1
\end{array}
\right)
\end{equation}

\noindent The secondary constraint is

\begin{eqnarray}
\label{formula20}
\Omega^{(2)} &=& v_\alpha^{(1)} \frac{\partial V^{(1)}}
{\partial A_\alpha^{(1)}} \nonumber \\ \nonumber \\
&=& a_i\pi_i \approx 0.
\end{eqnarray}

\noindent Here we must mention that the primary and secondary constraints, Eqs(\ref{formula15}) and (\ref{formula20}), respectively, derived
by the Faddeev-Jackiw procedure are the same obtained by the Dirac formalism. Taking the time derivative of this constraint and 
introducing the result into the Lagrangean (\ref{formula16}) 
by means of a Lagrange multiplier $\eta$, we get a new 
Lagrangean $L^{(2)}$

\begin{equation}
\label{formula21}
L^{(2)} = (\pi_i + \rho a_i + \eta \pi_i)\dot{a}_i 
+ \eta a_i \dot{\pi}_i - V^{(2)},
\end{equation}

\noindent where $V^{(2)}$ = $V^{(1)}$. The new sympletic enlarged 
variables are  $\xi_j=(a_j,\pi_j,\\ \rho,\eta)$. The new 
coefficients are

\begin{eqnarray}
\label{formula22}
A_{a_i}^{(2)} &=& \pi_i + \rho a_i + \eta\pi_i, \nonumber \\
A_{\pi_i}^{(2)} &=& \eta a_i, \nonumber \\
A_{\rho}^{(2)} &=& 0,\nonumber \\
A_{\eta}^{(2)} &=& 0. \nonumber
\end{eqnarray}

\noindent And the matrix $f^{(2)}$ read as

\begin{equation}
f^{(2)}=\left(
\begin{array}{cccc}
0           & -\delta_{ij}  & -a_i   &  -\pi_i \\
\delta_{ij} &         0     &   0    &    -a_i \\
a_i         &         0     &   0    &     0   \\
\pi_i       &        a_i    &   0    &     0  
\end{array}
\right)
\end{equation}

\noindent where rows and columns follow the order: $a_i$, $\pi_i$,$\rho$
 $\eta$. The matrix $f^{(2)}$ is not singular. Then we can identify 
it as the sympletic tensor of the constrained theory. The inverse 
of $f^{(2)}$ will give us the Dirac brackets of the
physical fields and can be obtained in a straightforward calculation. 
The resulting inverse matrix is

\begin{equation}
(f^{(2)})^{-1}=\left(
\begin{array}{cccc}
0                   &  \delta_{ij}-a_ia_j  &     a_i        &     0 \\
-\delta_{ij}+a_ia_j &  a_j\pi_i-a_i\pi_j   &    -\pi_i      &    a_i \\
-2a_i               &        -2\pi_i       &      0         &  -\delta_{ij}   \\
0                   &       2 a_i          &   \delta{ij}   &     0  
\end{array}
\right)
\end{equation}

\noindent From the matrix $(f^{(2)})^{-1}$ we identify the 
nonvanishing brackets

\begin{eqnarray}
\label{formula24 }
\{ a_i,a_j \}& = & 0,\\
\{ a_i,\pi_j \}& = & \delta_{ij} - a_i a_j ,\\
\{ \pi_i,\pi_j \}& = & a_j\pi_i-a_i\pi_j .
\end{eqnarray}

\noindent Then, by means of the well known canonical quantization rule
$\{\,,\,\}^* \rightarrow -i\[\,,\,\]$, we obtain the
commutators

\begin{eqnarray}
\label{formula25}
\[ a_i,a_j \]& = & 0,\\
\[ a_i,\pi_j \]& = & -i \( \delta_{ij} - a_i a_j \) ,\\
\[ \pi_i,\pi_j \]& = &  -i \( a_j\pi_i - a_i\pi_j \).
\end{eqnarray}

\noindent We observe that the brackets above, derived by Faddeev-Jackiw
Lagrangean procedure, are the same obtained by using the Dirac 
Hamiltonian theory.

Finalizing, when we treat the Born-Infeld skyrmions Lagrangean,  
approximating it by a finite second-order sum in the rotational 
modes variables, we verified that its quantum structure, given by
the commutators, are the same obtained in the conventional Skyrme 
model quantization. The expansion of the non-polynomial Born-Infeld 
Lagrangean in terms of dynamic variables is possible if we pay 
attention to the problem of breaking the relativistic invariance in 
the collective coordinates expansion. 
The non-causal soliton solution contributions to the physical parameters
are not relevant if the chiral angle F(r) satisfied the relation, 
$F(r) \rightarrow 0$, in the limit, $ r \rightarrow \infty $, together with 
the fact that we impose the soliton angular velocity small, i.e., 
$ \omega \ll 1$. Thus, taking the angular soliton solution, $\omega$, 
as an expansion parameter, and maintaining the non-causal soliton 
sector contributions as minimum as possible, we conclude that the quantum 
commutators of the Born-Infeld Skyrmions move towards to the 
conventional Skyrme ones.

This work is supported in part by FAPEMIG, Brazilian Research Council.

\end{document}